# Polystyrene grafting from silica nanoparticles via Nitroxide-Mediated-Polymerization (NMP): synthesis and SANS analysis with contrast variation method

Chloé Chevigny,[a] Didier Gigmes,[b] Denis Bertin,[b] Jacques Jestin,[a*] and François Boué[a]

We present a new convenient and efficient "grafting from" method to obtain well defined
10 polystyrene (PS) silica nanoparticles. The method, based on Nitroxide-Mediated Polymerization (NMP), consists to bind covalently the alkoxyamine, which acts as initiator controller agent, at the silica nanoparticles surface in two steps. The first step is a reaction between the aminopropylsilane and the silica particles in order to functionalize the particles surface with amino group. In a second step, the initiating-controlling alkoxyamine moiety is introduced via an over grafting reaction
15 between the amino group and the N-hydroxysuccinimide based MAMA-SG1 activated ester. To simplify both their chemical transformation and the polymerization step, the native silica particles, initially dispersed in water, have been transferred in an organic solvent, the dimethylacetamide, which is also a good solvent for the polystyrene. The synthesis parameters have been optimized for grafting density, conversion rates, and synthesis reproducibility while keeping the colloidal
20 stability and to avoid any aggregation of silica particles induced by the inter-particles interaction evolution during the synthesis. After synthesis, the final grafted objects have been purified and the non-grafted polymer chains formed in the solvent have been washed out by ultra filtration. Then the particles have been studied using Small angle Neutron Scattering (SANS) coupled to neutron contrast variation method. To optimize the contrast conditions, both hydrogenated and deuterated
25 monomers have been used for the synthesis. A refined fitting analysis based on the comparison on two models, a basic core-shell and the Gaussian Pedersen model, enables us to fit nicely the experimental data for both the hydrogenated and deuterated grafted case. Differences are seen between grafting of normal or deuterated chains which can be due to monomer reactivity or to neutron contrast effect variations. The synthesis and the characterization method established in this
30 work constitute a robust and reproducible way to design well defined grafted polymer nanoparticles. These objects will be incorporated in polymer matrices in a further step to create Nanocomposites for polymer reinforcement.

## Introduction

Creating model nano-composites composed of nano-fillers
35 introduced in a polymer matrix is a step towards the understanding of mechanical reinforcement of polymers, by studying the relation between the filler structure and the obtained mechanical properties[1]. Two different contributions to reinforcement are often distinguished: (i) the interaction of
40 the matrix polymer chains with the surface of the nanoparticles, (ii) the filler spatial diserpsion in the matrix.

[a] *Laboratoire Léon Brillouin, CEA Saclay 91191 Gif sur Yvette Cedex France*
45 [b] *Laboratoire Chimie Provence, UMR 6264, CNRS et Universités d'Aix-Marseille 1,2 et 3, Site de St Jérôme, Av. Esc. Normandie-Niemen case 542, 13393 Marseille Cedex 20, France*
[*] *corresponding autor : jacques.jestin@cea.fr*

50 The main dominant point which will permit to understand and describe such mechanisms is the control of the dispersion state of the filler in the polymer matrix. Such control could be based on an external trigger by a simple control of the film processing conditions[2], of the electrostatic repulsion[3] or with
55 a magnetic field[4] but also using an internal way based on advanced chemistry and particularly grafting techniques of polymer chains on the surface of the particles. The grafting can be performed "onto"[5] or "from" the surface of the particles using preferably various controlled polymerization
60 techniques[6] as ATRP (Atom Transfer Radical Polymerization)[7-15], NMP (Nitroxide Mediated Polymerization)[16-25] or RAFT (Reversible Addition Fragmentation chain Transfer)[26-27]. Synthesis of particles with chains grafted covalently at their surface should enable to
65 control, depending on the structure of the grafted chains layer, these two contributions (i) and (ii), and therefore to determine their dispersion. The synthetic method selected (polymerization techniques, degree of control, reproducibility) depends on the nature of the polymer, the polymerization medium (bulk, aqueous or organic solution) and the quantities of grafted particles needed for further applications. In our team, we were interested in both starting from a colloidal suspension of well dispersed nanoparticles in an organic solvent and maintain the stability and the good dispersion in the sol during grafting. Hence we would be able to start from individual nano-objects to introduce them in a polymer matrix: this final aim will be reported in a subsequent paper. In order to reach this goal, we were needing a reliable chemical synthesis route, with high conversion and good reproducibility to produce in large quantity, from an original silica sol, grafted nano-beads quite well defined from the beginning, and still well dispersed in the final sol after grafting. Moreover controlled polymerization is crucial to have a well defined corona in term of architecture, composition and distribution (to study later its effect on the mechanical properties). The complete and accurate characterization of the sol and of the objects after the synthesis will be then performed by means of Small Angle Neutron Scattering (SANS). This technique gives a deep characterization of the dispersion and of the structure of each object. Indeed thank to the annihilation of the signal either of the core or of the corona, it is possible to observe the spatial distribution of the other component. Moreover, it would be interesting to synthesize the corona out not only of normal polystyrene (h-PS), but also of deuterated polystyrene (d-PS); this would be useful when studying the dispersion of the grafted particles in a polymer matrix, as planned in the further work.

The first part of the paper is devoted to the description of the grafting-from method we developed to prepare the polystyrene silica nanoparticles. Among the different grafting-from controlled radical polymerization techniques already reported, we have chosen a process based on the NMP. Even if RAFT, ATRP or NMP systems have all proven their efficiency to elaborate polymer silica nanoparticles, compared to the RAFT and ATRP techniques, NMP is a mono-component system that is undoubtedly an advantage for such application. Indeed, during the synthesis, maintaining the stabilization of the colloidal suspension is a major concern. In this context we do believe that the use of a multiple component initiating system (e.g metal halide and ligand for ATRP, or conventional initiator in addition of the RAFT moiety attached to the silica nanoparticles) increases the source of possible aggregation phenomena by disturbing the particle interactions. Moreover, regarding the reinforcement applications unlike in NMP in the case of ATRP the polymer grafted silica nanoparticles have to be purified to get rid of the metal complexes prior to the nanocomposite formulation step. The interest of our approach is related to its simplicity, versatility and robustness. Indeed, while different strategies have been proposed to attach an alkoxyamine onto a silica particle, there is still a need for efficient and convenient initiator grafted procedure. Our approach consists to bind covalently the alkoxyamine, which acts as initiator controller agent, in two steps onto the silica nanoparticles surface. The first step is a reaction between the aminopropylsilane and the silica particles in order to functionalize the particle surface with amino group. In a second step the initiating-controlling alkoxyamine moiety is introduced via an overgrafting reaction between the amino group and the N-hydroxysuccinimide based MAMA-SG1 (MAMA-NHS) activated ester previously prepared in a straightforward manner from the commercially available MAMA-SG1 (blocbuilder) (figure 1)[28]. We have chosen the overgrafting method because the latter has been already used for the functionalization of silica wafers or silica particles and was found to give better results in terms of control and degree of grafting compared to other methods. We use an alkoxyamine bearing a tertiary stabilized alkyl moiety as initiating species, which gives a much better control of the polymerization reaction compare to the secondary alkyl stabilized one usaually employed[29].

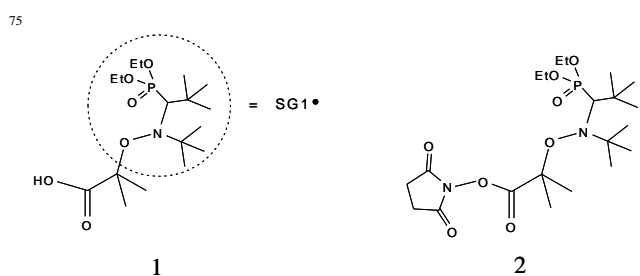

**Fig. 1** Initiator MAMA-SG1 (BlocBuilder), molecule 1, and activated ester MAMA-NHS, molecule 2.

The second part of the paper then describes the detailed characterization of the grafted nanoparticles successfully previously prepared, with a high degree of confidence, using SANS combined with variable contrast[30] between the two components (the silica core and the grafted polymer corona) of the grafted particles. Indeed, adding to the sol the same organic solvent but in its deuterated form, we can vary the deuterated fraction in order to observe either the silica or the polymer. If we annihilate the contrast between the solvent and the grafted chains, we can observe the shape and size distribution of the cores, and the spatial distribution of their centres of mass, and show how close they can be from the ones for the initial particles. Observations of the core after synthesis have been reported by several other groups[31, 20]. But in this paper, we also annihilate the contrast between solvent and cores, enabling us to observe the polymer shell around the cores if grafting has been successful. This method has been used many times in dispersions of various hybrid objects, including centro-symmetric ones like surfactants or copolymers micelles[32]. But in our knowledge it has been used only a few times for grafted nanoparticles. First some qualitative observations, which followed the grafted polymer growth during synthesis, were reported[8-9]. For a more quantitative and accurate description of the objects, we need to use a geometrical model representing our grafted objects. We discuss here which is the most convenient. A first attempt was reported using a core-shell model[7], while we also use here the Pedersen model[33-34] first developed for copolymer micelles. In this paper we have used in particular a new type of silica nanoparticles with a convenient size and a narrow enough size distribution; combined with the fact that the stability is well kept after synthesis, this enables a more thorough fitting. The whole process is achieved for corona synthesized either out of normal polystyrene or out of deuterated polystyrene.

## Experimental

### Material

Aminopropyltriethoxysilane (APTES, stored under nitrogen), N-hydroxysuccinimide, Dicyclohexylcarbodiimide, Styrene and the solvent dimethylacetamide (DMAc) were used as received from Aldrich. MAMA-SG1 (BlocBuilder®) was kindly provided by Arkema and used as received. Deuterated styrene was purchased from Eurisotop and used as received.

We use silica Ludox TM-40, purchased from Aldrich. It is a colloidal suspension of mono-disperse silica beads in water; they are transferred in DMAc by evaporation. The Ludox are first dissolved in a large volume of water, and the same volume of DMAc is added. The resulting solution is then heated to 100°C, under agitation, to evaporate water and concentrate the suspension; we stop when silica concentration is 5 % in weight, which also correspond to complete evaporation of water. The final suspension of Ludox silica in DMAc is characterized by Small-Angle Neutron Scattering (SANS) to check the dispersion (see results).

### Synthesis

#### Silanization of colloidal silica nanoparticles in DMAc

In a typical run, the silica dispersion (100 g of 5wt% $SiO_2$ in DMAc) is added to a 250 mL round-bottom flask with a magnetic stir bar. Aminopropyltriethoxysilane (APTES) (395 mg, 1.79 mmol, which corresponds to about 1 molecule/nm2) and methanol (6.6 mL) are added and the reaction mixture is left reacting for 24 hours at room temperature. The reaction mixture is then filtered under nitrogen pressure using a Millipore Ultra-filtration apparatus with a 30 000 Dalton pores diameter filter (regenerated cellulose) purchased from Millipore, to purify the solution from unreacted silanes. The solution is filtered four times. Each time, 100 mL of the obtained solution is diluted with 200 mL of DMAc and concentrated to the initial volume.

#### Over-grafting of the initiator

MAMA-NHS previously prepared according to[27] (268 mg, 0.56 mmol) is added to the silanized silica solution (100 g of 5wt% $SiO_2$ in DMAc). The reaction takes place during two hours, under nitrogen bubbling and at 0°C, to avoid the dissociation of the initiator. The solution of initiator-grafted particles is then filtered (same procedure as describe before) at 0°C. Thermo-gravimetric analyses were used to determine the yield of grafting and reaction and therefore the amount of initiator.

#### Polymerization from the surface of the nanoparticles

Model polymerizations in solution, in exactly the same conditions, but without silica beads, were first performed to check the feasibility and the good control of the polymerization. The solution of initiator-grafted silica particles is diluted to 1.1wt%, and 150g are added to a 500 mL three-neck flask. Styrene (50g) is added dropwise to the solution within the first hour, in order to avoid gelling, under constant stirring. Free initiator (BlocBuilder®, 339 mg, 0.89 mmol) is added to the solution to ensure a better control of the polymerization. The reaction mixture is then deoxygenated by nitrogen bubbling for 30 min. The flask is then put in an oil bath at 120°C to launch the polymerization. The reaction takes place for 6 hours, under nitrogen pressure. Kinetic samples are taken via sterile syringes and used to determine conversion by gravimetry.

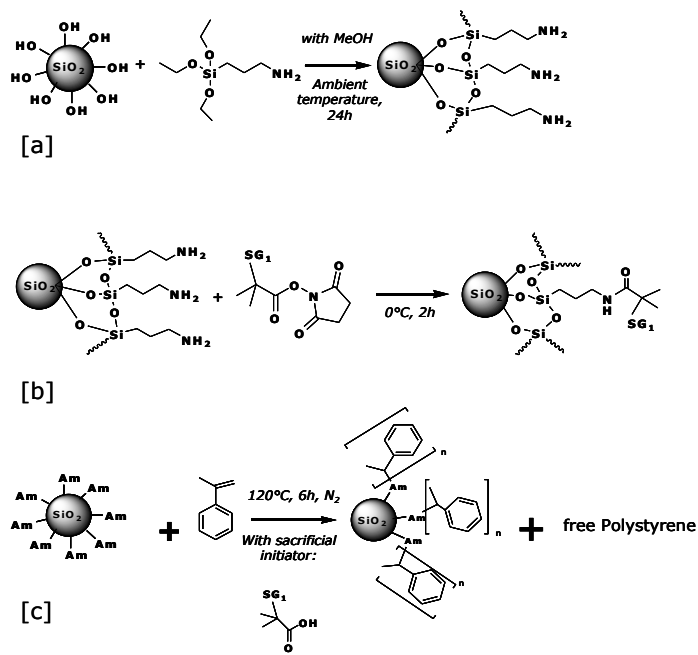

**Fig. 2** Scheme of the three steps of grafting: the silanization [a], the over-grafting of the initiator [b] and the polymerization from the particles surface [c].

When the polymerization is over, we have to separate the polystyrene-grafted particles from the free polystyrene in solution (from the free initiator). We use the same ultra filtration device as before, but with a 100 000 Dalton pores diameter filter in regenerated cellulose. Each time, 100 mL of the solution is diluted with 200 mL of DMAc and concentrated to the initial volume; the solution is filtered five times. We keep apart the free polymer chains, which are characterized by Size Exclusion Chromatography (SEC). The procedure is exactly the same for deuterated styrene ($C_8D_8$) and for styrene $C_8H_8$. After extensive characterization of the sol by TGA to determine grafting density of the polymer, the solutions are prepared at the desired concentration for SANS measurements. Figure 2 resumes the whole grafting process.

**Characterization**

The grafting densities (silane, initiator and polymer) are determined by Thermo-gravimetric analysis (TGA). We use a TA instrument Q50, at a scan rate of 10 °C.min$^{-1}$ to 800 °C, under a nitrogen flow (60 mL.min$^{-1}$); the grafting densities are determined using the equation below:

$$Grafting\ density = \frac{S_{spe}}{M_{gr} * Na} * \frac{W_{tot} - W_{ref}}{100 - (W_{tot} - W_{ref})} \quad (1)$$

Where $S_{spe}$ is the specific surface (nm$^2$/g) of the silica, $M_{gr}$ the molar mass of the grafted molecule, $N_a$ the Avogadro number, $W_{tot}$ and $W_{ref}$ are, respectively, the weight loss of the grafted sample and the weight loss of the reference sample (silica for determining the initiator grafting and initiator-grafted silica for the polymer).

**Small Angles Neutron Scattering (SANS)**

**Measurements**

Measurements were performed at the Laboratoire Léon Brillouin (LLB) on the SANS spectrometer called PAXY. Three configurations were used: the first one with wavelength 15 Å, sample-to-detector distance of 6.70 m, and a collimation distance of 5.00 m, and the second with wavelength 6 Å, sample-to-detector distance of 6.70m, and a collimation distance of 2.50 m and the last one with wavelength 6 Å, sample-to-detector distance of 3.00m, and a collimation distance of 2.50 m corresponding to a total Q range of 2.10$^{-3}$ Å$^{-1}$ to 0.1 Å$^{-1}$. Data processing was performed with a homemade program following standard procedures with $H_2O$ as calibration standard. Small deviations, found in the spectra at the overlap of two configurations, are due to different resolution conditions and (slight) remaining contributions of inelastic, incoherent, and multiple scattering. To get the cross-section per volume in absolute units (cm$^{-1}$), the incoherent scattering cross section of $H_2O$ was used as a calibration. It was estimated from a measurement of the attenuator strength, and of the direct beam with the same attenuator. The incoherent scattering background, mainly due to protons of the solvent, was subtracted using a blank sample with zero silica fractions.

**Neutrons Contrast variation method**

The contrast variation method offered by neutron scattering experiments is a powerful technique to elucidate complex structures made out of the association of binary component systems, i.e. here the polymer chain, the silica particle and the solvent. Two main conditions must be realized: the scattering length density of the two components of the system must be sufficiently different from each other and these values must be also comprised between the scattering length density values of the hydrogenated and the deuterated solvent. This is the case for our system regarding the following scattering length density ($\rho$) values: $\rho_{SiO2}$ = 3.40 10$^6$ Å, $\rho_{h-PS}$ = 1.43 10$^{-6}$ Å$^{-2}$, $\rho_{d-PS}$ = 6.53 10$^{-6}$ Å$^{-2}$, $\rho_{h-DMAc}$ = 0.52 10$^{-6}$ Å$^{-2}$, $\rho_{d-DMAc}$ = 6.60 10$^{-6}$ Å$^{-2}$. The total scattering length density of a mixture of hydrogenated and deuterated solvent can be written as a function of the respective proportion of H/D in volume fraction $\Phi$ as:

$$\rho_{mixture} = \rho_{h-DMAc} \times \Phi_{h-DMAc} + (1-\Phi_{h-DMAc}) \times \rho_{d-DMAc} \quad (2)$$

Using this property, we can perform SANS measurements on a solution of grafted silica nanoparticles in contrast matching condition for which we can first match the scattering contribution of the polymer chain to check the colloidal stability of the suspension. This can be done with a mixture of 85%/15% h-DMAc/d-DMAc. Secondly, on the same sample, we can investigate the scattering contribution of the grafted PS corona by matching the silica core contribution with a mixture of 53%/47% h-DMAc/d-DMAc. The resulting scattering signal in both contrast cases can then be analysed by comparison to calculated scattering, using some specific expressions of form factor that we described just below.

**Fitting Models**

The total intensity I(Q) scattered by a colloidal solution of centrosymetrical particles volume fraction $\Phi$ can be written as follows:

$$I(Q) = \Phi \rho^2 V_{part} F(Q) S(Q) \quad (3)$$

$V_{part}$ is the volume of the particle, $\Delta\rho^2$ is the difference between the scattering length density of the particle and the scattering length density of the solvent, F(Q) is the from factor of the particle and S(Q) is the structure factor of the particles. For diluted colloidal solutions, the interactions between the particles can be ignored and we can consider the structure factor to be close to 1 (S(Q)~1). The form factor of a bead which we assume to be a compact sphere is written as:

$$F(Q)^2 = [3 f(QR)]^2 \quad with \quad f(x) = \frac{\sin(x) - x\cos(x)}{x^3} \quad (4)$$

R is the radius of a native particle. The polydispersity in size of the silica beads is described by means of a log-normal distribution with parameters $R_0$ and $\sigma$.

$$P(R, R_0, \sigma) = \frac{1}{\sqrt{2\pi} R \sigma} \exp\left(\frac{-1}{2\sigma^2} \ln^2 \frac{R}{R_0}\right) \quad (5)$$

The form factor with polydispersity is calculated by integration:

$$F^2(Q) = \int P(R_{SiO_2}, R_0, \sigma) F(Q, R_{SiO_2}) dR_{SiO_2} \quad (6)$$

This form factor can be used to analyze the scattering signal of the native silica particles and also to analyze the scattering of the grafted silica particles in contrast matching conditions for which the polymer is matched, i.e. we see only the silica core. For the other contrast matching condition in which we match the silica particle to see only the grafted polymer chains, two main form factors will be considered. The first one, and the most basic, is the core shell model which can be expressed as:

$$F_{shell}^2(Q, R_{SiO_2}) = \left[ \begin{array}{c} \frac{4\pi}{3}(R_{SiO_2} + e)^3 f(Q(R_{SiO_2} + e)) \\ -\frac{4\pi}{3} R_{SiO_2}^3 f(QR_{SiO_2}) \end{array} \right]^2 \quad (7)$$

f(x) is defined in equation 3, the polydispersity of the radius of silica beads could be included by integration like equation 5, and $e$ denotes the thickness of the shell. This model supposes that the corona is made of constant and homogenous density, which can be a mixture of polymer and solvent of constant concentration, as function of the distance from the interface and will be consequently suitable for high grafting density. The second model we use has been built by Pedersen[33] for block copolymer micelles and is representative of the form factor for Non-interacting Gaussian chains. The general expression can be written as follows:

$$F^2_{pedersen}(Q, R_{SiO_2}) = V_{core}^2 \Delta\rho_{core}^2 F_{core}(Q) + N\Delta\rho_{chain}^2 V_{chain}^2 F_{chain}(Q)$$
$$+ 2NV_{core} V_{chain} \Delta\rho_{core} \Delta\rho_{chain} S_{core-chain}(Q) +$$
$$N(N-1)\Delta\rho_{chain}^2 V_{chain}^2 S_{chain-chain}(Q)$$
$$(8)$$

Where $V_{core}$, $V_{chain}$ are the volume of the core and of the chain, $F_{core}$, $F_{chain}$ are the form factors of the core and of the chain, $\Delta\rho^2_{core}$, $\Delta\rho^2_{chain}$, are the contrast between the core and the solvent and between the chain and the solvent, $S_{core-chain}$ and $S_{chain-chain}$ are the structure factor between the core and the chain and the inter chain structure factor, N is the number of chains. The form factor of the core $F_{core}$ can be written as a sphere form factor (equation 3). The form factor of the Gaussian chain $F_{chain}$ can be written according to the classical Debye expression:

$$F^2_{chain}(Q) = \frac{2(e^{-x} - 1 + X)}{X^2}, \text{ with } X = Q^2 R_g^2 \quad (9)$$

where $R_g$ is the radius of gyration of the Gaussian chain. The crossing term related to the interaction between the core and the chain can be expressed as:

$$S_{core-chain}(Q) = \frac{3}{Q^3}\left[\frac{\sin(QR) - (QR)\cos(QR)}{R^3}\right] \times$$
$$\frac{[1 - \exp(-Q^2 R_g^2)]}{(Q^2 R_g^2)} \frac{\sin[Q(R_g + R)]}{[Q(R_g + R)]} \quad (10)$$

The intra chain form factor can be expressed as:

$$S_{chain-chain}(Q) = \frac{[1 - \exp(-Q^2 R_g^2)]}{(Q^2 R_g^2)^2} \left(\frac{\sin[Q(R_g + R)]}{[Q(R_g + R)]}\right)^2 \quad (11)$$

In contrast matching conditions for which the silica core is matched to see only the polymer corona, the Pedersen expression (equation 7) is reduced to:

$$F^2_{pedersen}(Q, R_{SiO_2}) = N\Delta\rho_{chain}^2 V_{chain}^2 \times$$
$$[F_{chain}(Q) + (N-1)S_{chain-chain}(Q)] \quad (12)$$

And the total measured intensity I(Q) as:

$$I(Q) = \frac{\Phi}{V_{part}}\left[N\Delta\rho_{chain}^2 V_{chain}^2 [F_{chain}(Q) + (N-1)S_{chain-chain}(Q)]\right] \quad (13)$$

In which the polydispersity can be included by integration of this expression according to the equation 5. In the case of block co-polymer, this model must contain an additional parameter illustrating the inter-diffusion zone between the two polymers and the non possibility for the chain to penetrate into the polymer core must strictly be imposed by choosing an appropriate radial profile for the corona[34]. It is not necessary in our case as it is physically impossible for the grafted chains to penetrate the silica core and thus simplified the analytical formulation of the model. The expressions of form factor for interacting self-avoiding chains have been derived by Svaneborg and Pedersen on the basement on Monte Carlo simulations[35]. We do not try to analyse our data with this model as the interactions between the chains in the corona can be neglected regarding the range of the experimental grafting density. To analyze the experimental data, we have developed a home made code under Matlab with a fitting procedure using a least-square regression method. For each form factor model describe below, a routine procedure have been writing including the fitting parameters. The numbers of fitting parameters is model dependent and can be reduced by fixing them to calculated (contrast term) or experimental (silica volume fraction) values when it is possible. The routine is running with various combinations of the adjustable fitting parameters for which a fitting criterion, $\chi^2$, is calculated. All parameters are restricted within a range of values of physical meaning. We repeat the process until obtaining a minimal value of $\chi^2$.

## Results

### Native silica particles

The control of the aggregation state of the silica Ludox particles after the solvent transfer is checked by comparison of the SANS spectra of a dilute solution of the particle in water (figure 1 [a]) and in Dmac (figure 1 [b]).

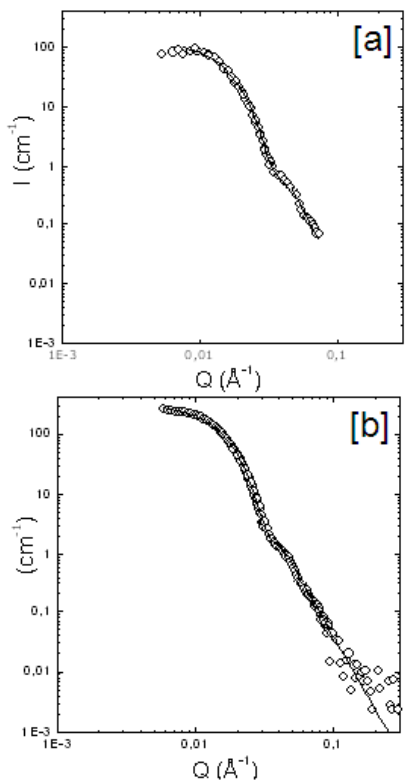

**Fig. 3** Scattering curves of diluted solution of Ludox Silica particles in water [a], and in Dmac [b] after solvent transfer. Open circles are experimental data and full line are fit with a polydisperse form factor of spheres.

The scattering curves have been analyzed using the fitting routine corresponding to a polydisperse form factor of spheres (equation 4 and 6, and fitting procedure described in the material and methods section). This model contains four parameters: the silica volume fraction $\Phi$, the contrast term $\Delta\rho^2$, the mean sphere radius R and the polydispersity $\sigma$. The contrast term can be calculated and is fixed to reduce the number of fitting parameters ($\Phi$, R, $\sigma$). The scattering curve of the initial solution of Ludox particles in water is nicely fitted by this model which gives a best fit value for the mean sphere radius equals to 134 Å with a polydispersity equal to 0.18. The solution of the same Ludox silica particles after transferred in the DMAc solvent is also well fitted with the similar polydisperse form factor indicating that the particles are well disperse in the organic solvent, i.e., there is no aggregation due to the change of solvent. It seems us that such stability must be governed by the polarity of the DMAc but we do not discuss more precisely this point since this is currently not the scoop of this paper. The very important result for us is that using a simple solvent transfer procedure, we are able to obtain well-defined silica particles in a well-dispersed state and in a solvent in which the NMP polymerization of the polystyrene is efficient.

### Polymerization, chemical characterization and purification

We need to obtain good conversion rates for the polymerization while keeping the control of the colloidal stability during the synthesis to avoid aggregation of the particles. Firstly, we have defined the conditions of synthesis using a model polymerization, without silica particles, to verify the control of the polymerization using the MAMA-NHS and to have an estimate of the conversion rates. The result of this model polymerization (120°C, $M_{n\ theoretical}$ 50 000 g/mol and 30% v/v of monomer) of Polystyrene in DMAc is reported on figure 4. Chain molecular weight distribution of the free chain displays a low polydispersity index, less than 1.3, through the whole polymerization.

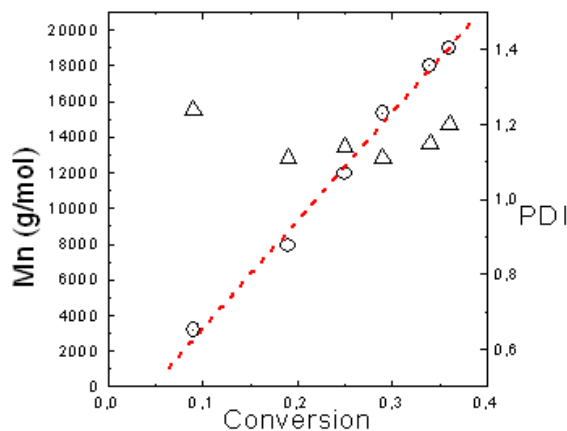

**Fig. 4** Evolution of the molecular masse Mw (left, open circles) and polydispersity index PDI (right, open triangles) as function of the conversion for the model polymerization of polystyrene with NMP polymerization. The red line is a guide for the eyes illustrating the control of the polymerization.

The number average molecular weight $M_n$ varies linearly with conversion (up to 50%), close to the theoretical line as expected for a controlled polymerization with an efficient initiation. Initially we were interested in changing the molecular weight of the grafted chains by varying the synthesis conditions. This can be done by varying the ratio $\frac{n_{initiator}}{n_{monomer}}$, which corresponds to the theoretical $M_n$ of the polymer chains at 100% of conversion. Some tests in this direction on model polymerizations reveal that reducing the initiator concentration reduces too much the efficiency and particularly the control of the polymerization. The other way to increase the theoretical length of the molecular chain is to increase the monomer concentration. This is also not convenient because introducing more styrene in the system changes drastically the polarity of the media resulting in the destabilization of the sol. Some preliminary tests showed us that, for a dilute solution of silica particles (typically 1% v/v), we cannot increase the monomer concentration up to the one deduced from the model polymerization, typically 30% v/v. After the transfer from water to DMAc as detailed previously, the silica particles were thus first covered with Aminopropyltriethoxysilane and purified from free silane molecules by ultra filtration. The content of silane and silica in the resulting dispersion has been characterized by TGA analysis. Two different batches of silanised silica, entries 1 and 2 have been successively synthesized within the same conditions. Results are presented in table 1. The grafting densities of silanes are relatively high and we can note some small differences between the two batches 1 and 2, but the values are nevertheless close enough from each other to be directly comparable. The following step of polymerization, the over-grafting of the initiator (MAMA-NHS), has been conducted also on these two batches. After purification by ultra filtration, the content of grafted organic material has also been characterized by TGA (see table 1) on the two over grafted silica batches.

**Table 1** Chemical analysis deduced from thermo gravimetric measurements on the four polymerizations.

| Polymerization | Silane/nm² | Initiator/nm² | Chain/nm² | Chain/Part. | Mn (g/mol) | Mw (g/mol) | PDI | Conversion Rate (%) | Initiation efficiency (%) |
|---|---|---|---|---|---|---|---|---|---|
| $h_1$ | 0.48 | 0.19 | 0.14 | 382 | 24000 | 32700 | 1.36 | 46 | 74 |
| $d_1$ | 0.48 | 0.19 | 0.15 | 411 | 29300 | 37900 | 1.30 | 42 | 79 |
| $h_2$ | 0.43 | 0.23 | 0.15 | 444 | 29300 | 35800 | 1.22 | 49 | 65 |
| $d_2$ | 0.43 | 0.23 | 0.18 | 489 | 24400 | 31000 | 1.27 | 30 | 78 |

In the final step, the polymerization was achieved on both batches 1 and 2 using the synthesis conditions deduced from the model polymerization: 1% v/v of silica particles, 30% of monomer, and the corresponding quantity of initiator for $M_{n\ theoretical} = 50\ 000$ g/mol, at 120°C during 6 hours. For the batch number 1, two polymerizations were performed, one with hydrogenated styrene (called $h_1$) and the other with deuterated styrene (called $d_1$), in similar concentration conditions (30% v/v). For the second batch (number 2) two polymerizations were also made. The polymerization using the hydrogenated monomer h-styrene (called $h_2$) was done in similar conditions than for $h_1$ and $d_1$. The second one using the deuterated monomer d-styrene (called $d_2$) was realized with a lower monomer concentration equal to 18% v/v because of a lack of deuterated styrene at this time. The results of the evolution of the conversion as a function of time for all polymerizations ($h_1$, $d_1$, $h_2$, $d_2$) are reported in figure 5. For comparison, the model polymerization is also reported. The representation used, $\ln([M]_0/[M])$ as a function of $t^{2/3}$, exhibits for each case a quasi-linear variation indicating a good control and efficiency of the polymerization whatever the batch (1 and 2), and whatever the nature of the monomer (h or d). It shows that chain growth (propagation) is as fast as initiation. We can observe a nice reproducibility ($h_1$, $d_1$ and $h_2$) for the polymerizations which were synthesized in similar conditions. The final conversion rates are reported in table 1. We can observe a small decrease of the conversion rate in presence of particles as compared to the model polymerization; we also note very similar values for $h_1$, $d_1$ and $h_2$ and a lower value for the $d_2$ polymerization in agreement with the lower value which was taken for the initial monomer concentration. For any polymerization, "sacrificial" free initiator is necessary to initiate the polymerization and to ensure a good control of the chain growth; without the free initiator, the SG1 concentration would be to low to attain the persistent radical effect, thus allowing control. The consequence is the presence of free (non-grafted) polymer chains in the solvent at the end of the synthesis. The grafted nanoparticles can be separated from the free polymer chains by an ultra filtration process. The grafted nanoparticles were analyzed by TGA to evaluate the number of grafted chains per particles. The free polymer chains were analyzed by SEC to determine the molecular weight of the synthesized chains. Previous studies[11, 15, 19, 24, 36] shows that the mass of the grafted chains are usually of the same order of magnitude as the molecular of the non grafted chains. Results are presented in table 1.

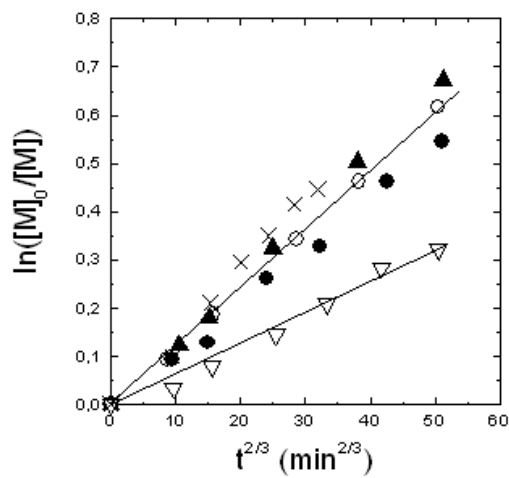

**Fig. 5** Semi logarithmic evolution of the conversion as a function time to the power 2/3 for the whole polymerization: the model (cross), the hydrogenated monomer $h_1$ (open circles), the hydrogenated $h_2$ (full circles), the deuterated monomer $d_1$ (full triangles) and the deuterated $d_2$ (open triangles).

**SANS results**

After purification, the different batches of grafted silica particles were measured with SANS measurements in dilute conditions (typically around 1% v/v) in both contrast matching conditions.

**Grafted particles in polymer matching condition**

According to the Scattering Length Density values of the components, the matching of the deuterated grafted chains could be achieved with a mixture 1/99 % v/v h-DMAc/d-DMAc. The latter contrast condition implies to exchange all but 1% of the non deuterated DMAc with some deuterated one. A simple dilution by 100 would give too much dilute samples for sufficient scattering. Dialysis would be possible, but would involve large quantities of deuterated solvent, which is costly. For these reasons, we only analyzed the contrast condition corresponding to the matching of the hydrogenated chains. In this case, the contribution of the hydrogenated polymer can be matched with a mixture of hydrogenated and deuterated DMAc equal to 85/15 % v/v. The Deuterated DMAc is added in convenient proportions in the purified solution of the polymerization $h_1$ and put in a quartz Hellma cell of thickness 1mm (due to the high content of protons from the solvent). The particle concentration of the solution (close to 1% v/v) was adjusted to reduce strongly the interaction between the particles and to measure only the particles form factor. The background of the sample, including mainly incoherent scattering, is determined by measuring the pure solvent, the mixture h-DMAc/d-DMAc 85/15 % v/v, in the same conditions. The corresponding flat signal is subtracted from the one of the particle solution as a function of the silica particles concentration $\Phi$ as:

$$I(Q) = I_{measured} - \left[ (1-\Phi) \times I_{solvent} \right] \quad (14)$$

The resulting signal, i.e. the scattering from hydrogenated polystyrene (h-PS) grafted particles for which the grafted chains are matched is reported on the figure 4. In the low-Q range of the spectra, the curve exhibits a plateau indicating a finite size of the scattering objects and no aggregation of the particles which is the most important information here. In the intermediate Q-range, we see an oscillation and the intensity decreases quickly as function of $Q^{-4}$ in the large Q-range. Such variation is mostly encountered in presence of sharp interfaces between a compact object and the external medium, which suggest strongly that the signal is effectively due to the silica particles with no contribution from the polymer corona. The experimental curve was analyzed using the fitting procedure described previously with the following adjustable parameters: the silica volume fraction $\Phi$, the mean radius of the spheres R and the polydispersity $\sigma$. The best-fitted calculation is plotted on top of data on the figure 6 [a] (full red line). We can observe a very good agreement between the calculation and the experimental data for $\Phi = 0.5\%$, R = 134 Å and $\sigma = 0.18$. So when we match the grafted polymer chains on the silica particles, the obtained scattering signal can be fitted with the same polydisperse form factor than the one used for the silica particle before the polymerization process (figure 3[b]). This shows that the polymerization induced no aggregation of the particles and that the chain growth is occurring from the surface of a single particle and not of a cluster of many native particles as in this previous work[8]. The inter-particles structure factor can be extracted by divided the total intensity with the calculated form factor; result is presented in the insert of the figure 6 [b]. As we can see, the obtained structure factor present a nicely oscillation characteristic of a repulsive liquid-like order of the particles in the solution. This indicate the absence of aggregation between the silica core. Due to low concentration, the range of the interactions is enough long to produce an inter-particle structure factor close to one.

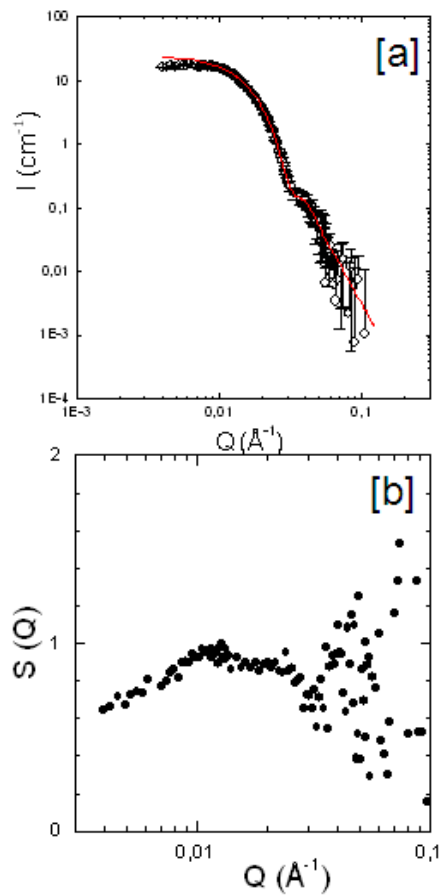

**Fig. 6** [a] Scattering curve of a dilute solution of grafted silica particles with hydrogenated PS chains in contrast condition of matching the scattering signal of the PS chains. Open circles are experimental data and the full red line is the best fitted result of the adjustment with a polydisperse form factor of spheres, [b] inter-particle structure factor deduced from the division of the total intensity (experimental data) by the calculated form factor (red curve).

## Grafted particles in silica matching condition: hydrogenated grafted chains

To see the corona only, the Scattering Length Density of the silica core on the polymer grafted particles can be matched with a solvent mixture h-DMAc/d-DMAc equal to 53/47 % v/v. We focus first on the dispersions after polymerizations of hydrogenated styrene ($h_1$ and $h_2$). Deuterated DMAc is added on diluted purified solution of grafted polymer particles in right proportions and put in quartz Hellma cells. The concentration of the particles has been adjusted to reduce the interactions between the objects inducing a structure factor close to unity. As the incoherent contribution is, in this contrast condition, lower due to the higher proportion of deuterated species, we can use cells of optical path of 2 mm to increase the coherent scattering signal while having an acceptable (though not very high) background level. This background is determined by measuring the pure solvent mixture h-DMAc/d-DMAc 53/47% v/v in the same experimental conditions. The resulting flat contribution is subtracted from the coherent one according to the silica particle concentration as in equation 14. The resulting signal of the hydrogenated PS grafted corona is reported on figure 7 for the polymerization $h_1$ and on figure 8 for the polymerization $h_2$. Let us first comment qualitatively the shape of the curves. In the low Q region, we can observe a flat plateau of the scattered intensity. The inter-particles structure factor, determined previously in the polymer contrast matching condition, has been found to be close to one (see figure 6 [b]).

This is the most robust way to check that there is no effective interaction between the particles and to be in the best conditions of the from factor determination. We can thus assume that this plateau correspond to finite size objects. Its extrapolation to the zero Q limit should give the total mass of the chains in the corona. This is a very safe information, independent of the fitting models. In the intermediate Q-range, we can see a nice oscillation followed by a decrease of the intensity in $Q^{-4}$. This $Q^{-4}$ behavior is followed in the large Q range by a decrease of the intensity as function of $Q^{-2}$. The transition between the $Q^{-4}$ and the $Q^{-2}$ regime is illustrated in insert (figure 8[b]) with a $Q^{-4}.I$ versus Q representation. The oscillation is related to the thickness of grafted polymer layer and the presence of the $Q^{-2}$ decrease is the signature of Gaussian chain conformation as expected if the grafted chains are swollen by DMAc which is a $\Theta$ solvent.

For each polymerization, we used the same approach of fitting first the experimental data with a core-shell model (equation 7) and secondly with the Gaussian chain model (equation 13). The core-shell model is composed of seven adjustable parameters: the particle concentration $\Phi$, the radius of core R, the polydispersity $\sigma$, the thickness of the grafted polymer layer e, the scattering length density of the silica core $\rho_{core}$, the one of the solvent $\rho_{solvent}$ and the scattering length density of the grafted corona $\rho_{corona}$. In this contrast condition, the scattering length of the solvent is equal and fixed to the scattering length density of the silica core: $\rho_{solvent} = \rho_{core}$. The scattering length density of the corona is range between the value of the polymer (a pure polymer grafted corona) and the one of the solvent (absence of grafting). In a first step fitting process, the radius R and the polydispersity of the silica core can be fixed to the values deduced from the previous contrast condition (figure 6) in which we see only the silica particles (R=134 Å, $\sigma$=0.18). This way permits to reduce the fitting parameters to only three ($\Phi$, e, $\rho_{corona}$) and thus to limit the risk of the creation of local minima of the fitting criteria and to obtain values without physical significations.

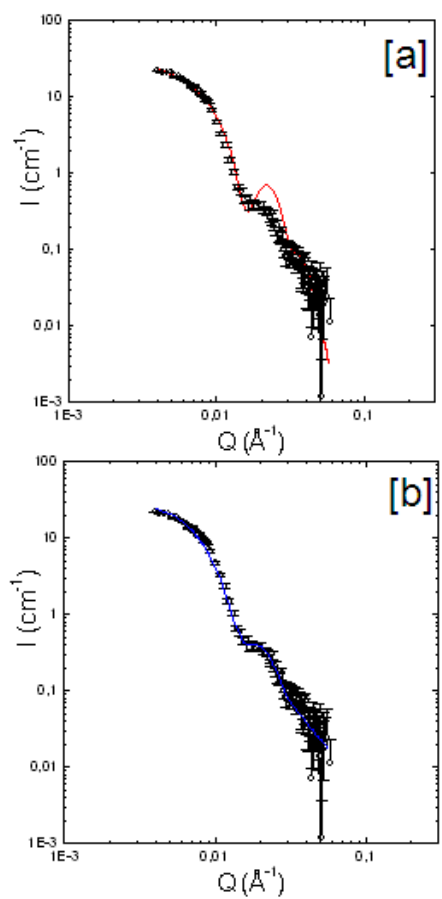

**Fig. 7** Scattering curve of a diluted solution of silica Ludox particles grafted with hydrogenated PS chains ($h_1$) in contrast conditions for which the scattering contribution of the silica core is matched with a mixture of hydrogenated and deuterated solvent (DMAc). Open circles are experimental data points. The full red line [a] is the best fitted result of the adjustment with a core-shell model and the full blue line [b] is the best fitted result of the adjustment with the Gaussian chain model.

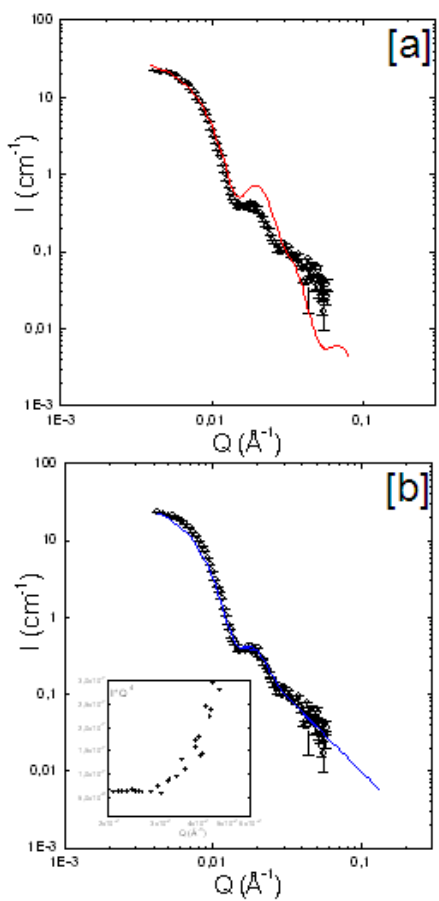

**Fig. 8** Scattering curve of a diluted solution of silica Ludox particles grafted with hydrogenated PS chains ($h_1$) in contrast conditions for which the scattering contribution of the silica core is matched with a mixture of hydrogenated and deuterated solvent (DMAc). Open circles are experimental data points. The full red line [a] is the best fitted result of the adjustment with a core-shell model and the full blue line [b] is the best fitted result of the adjustment with the Gaussian chain model.

**Table 2** Best-fitted results parameters of the fitting procedure applied on the hydrogenated polymerizations $h_1$ and $h_2$ with the two analyzed model, the core-shell and the Gaussian chain.

| Model | Polymerization | Concentration Φ% v/v | $R_{core}$ (Å) | σ Polydispersity | Shell Thickness | Shell Polymer volume fraction | Grafted Chains N | $R_g$ of grafted chains |
|---|---|---|---|---|---|---|---|---|
| Core-shell | $h_1$ | 0.50 | 134 | 0.18 | 98 | 0.31 | - | - |
| Core-shell | $h_2$ | 0.61 | 134 | 0.23 | 118 | 0.54 | - | - |
| Gaussian chain | $d_1$ | 0.70 | 130 | 0.25 | - | - | 360 | 58 |
| Gaussian chain | $d_2$ | 0.61 | 134 | 0.23 | - | - | 485 | 67 |

We first use the software to find a minimum of $\chi^2$ as function of the various combinations of the parameters (Φ, e, $\rho_{corona}$). The fit can be then refined by letting free the core parameters R and σ also. The same procedure is applied for the Gaussian chain model which is also composed of six parameters: the particle concentration Φ, the contrast term $\Delta\rho_{chain}^2$, the radius of core R, the polydispersity σ, the radius of gyration of the grafted chain $R_g$ and the number of the grafted chains N. The best fitting curve is the full red line for the core-shell model on figure 7[a] for the polymerization $h_1$ and on figure 8[a] for polymerization $h_2$. It is the full blue line for the Gaussian chain on figure 7[b] for $h_1$ and 8[b] for $h_2$. The best fitted parameters deduced from the fitting procedure are reported in table 2. The core-shell model reproduces the experimental data only for the low Q data for which it gives a good estimate of the particle volume fraction, of the global size and molecular mass of the grafted particles (core + polymer corona). The parameters of the core, R and σ, are very consistent with the one deduced form the scattering of the silica particles in solution (figure 6). There is also a good agreement with the chemical analysis results: the polymerization $h_2$ gives a higher conversion rate and larger chains than the polymerization $h_1$. This difference can be explained by the fact that the monomer $h_2$ have been distilled before the polymerization while it was not the case for $h_1$. This is traduced in the model by a larger thickness of the grafted corona and a higher polymer density in the corona. Nevertheless, the intermediate Q range, the oscillation, and the high Q domain (where the fit decreases very strongly compared to the data) are not well reproduced by the core-shell model.

On the contrary to the core-shell model, the results obtained with the Gaussian chain model fit much better the experimental data along the whole Q domain. The particle concentration and the parameter for the core are correctly described. The radiuses of gyration of the grafted chains are very consistent with the molecular masses deduced from the GPC experiments. Applying the well-know relation[37] for polystyrene, $R_g = 0.275 \cdot Mw^{0.5}$, we should expect to obtain values of the order of magnitude of 50Å, comparable to what we find, 58Å and 67Å. Finally, the number of grafted chains deduced from the fit is also consistent with the one deduced from the chemical analysis. To summarize, among two different models using the same number of adjustable parameters, the grafted hydrogenated PS particles are better described by the Gaussian chain than by the core-shell model.

**Grafted particles in silica matching condition: deuterated grafted chains**

Measurements and fitting procedures for the polymerization $d_1$ and $d_2$ have been realized in the same way as described for hydrogenated ones. The results are presented in figure 9 for the polymerization $d_1$ and in figure 10 for $d_2$. The shape of the curve is very similar to the one obtained for grafted chains $h_1$ and $h_2$ but we can note a difference which is illustrated in insert (figure 10[b]) in the $I.Q^4$ versus Q representation: the transition between the $Q^{-4}$ and the $Q^{-2}$ decrease of the intensity is shifted toward the larger Q value than for $h_1$ and $h_2$, typically from $2.8\ 10^{-2}$ to $3.3\ 10^{-2}$ Å$^{-1}$. In other words the Q-4 regime "survives" on a larger range before to "sink" below the $Q^{-2}$ scattering. The best fitting curve is the full red line for the core-shell model on figure 9[a] for the polymerization $d_1$ and on figure 10[a] for polymerization $d_2$, the full blue line for the Gaussian chain model on figure 9[b] for $d_1$ and 10[b] for $d_2$. The best fit parameters are reported in table 3. For each case, the parameters for the core (R, σ) are the one expected from the scattering from the naked particles and the particles concentration, Φ, is the experimental nominal one. The parameters for the Gaussian chain fit are consistent with the chemical analysis: the number of grafted chains and the radius of gyration of the chains are larger for the d1 polymerization.

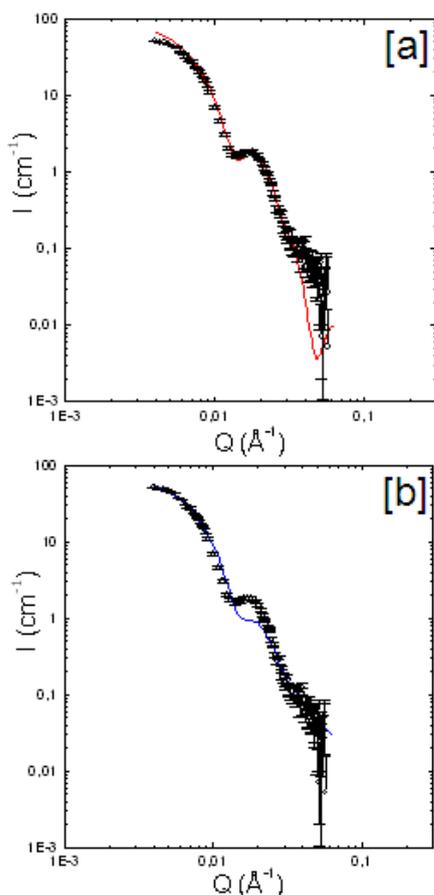

**Fig. 9** Scattering curve of a diluted solution of silica Ludox particles grafted with deuterated PS chains ($d_1$) in contrast conditions for which the scattering contribution of the silica core is matched with a mixture of hydrogenated and deuterated solvent (DMAc). Open circles are experimental data points. The full red line [a] is the best fitted result of the adjustment with a core-shell model and the full blue line [b] is the best fitted result of the adjustment with the Gaussian chain model.

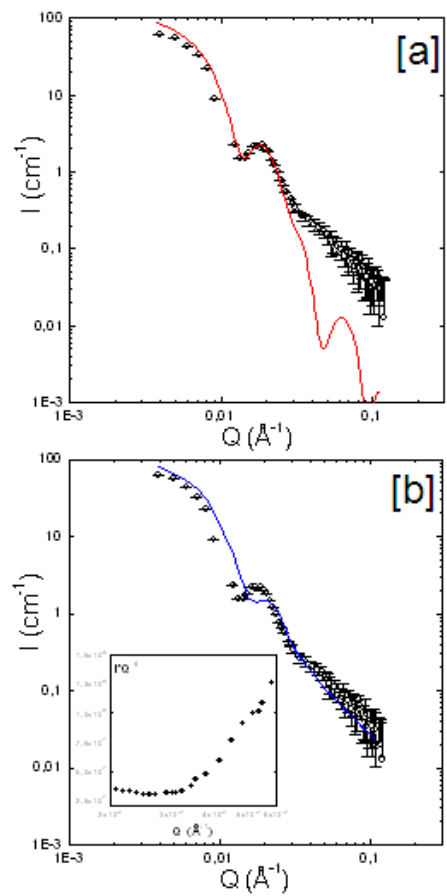

**Fig. 10** Scattering curve of a diluted solution of silica Ludox particles grafted with deuterated PS chains ($d_2$) in contrast conditions for which the scattering contribution of the silica core is matched with a mixture of hydrogenated and deuterated solvent (Dmac). Open circles are experimental data points. The insert is a $Q^4.I$ versus Q representation illustrating the cross-over between the $Q^{-4}$ and the $Q^{-2}$ regime. The full red line [a] is the best fitted result of the adjustment with a core-shell model and the full blue line [b] is the best fitted result of the adjustment with the Gaussian chain model.

**Table 3** Best-fitted results parameters of the fitting procedure applied on the deuterated polymerizations $d_1$ and $d_2$ with the two analyzed model, the core-shell and the Gaussian chain.

| Model | Polymerization | Concentration Φ%v/v | $R_{core}$(Å) | σ Polydispersity (Å) | Shell Thickness | Shell Polymer volume fraction | Grafted Chains N | $R_g$ of grafted chains |
|---|---|---|---|---|---|---|---|---|
| Core-shell | $d_1$ | 0.52 | 134 | 0.26 | 135 | 0.44 | - | - |
| | $d_2$ | 1.05 | 134 | 0.24 | 136 | 0.36 | - | - |
| Gaussian chain | $d_1$ | 0.52 | 134 | 0.26 | - | - | 410 | 57 |
| | $d_2$ | 1.05 | 134 | 0.24 | - | - | 320 | 48 |

This is consistent with the larger value of the conversion rate obtained for $d_1$ (42%) in comparison to the one of $d_2$ (30%). Nevertheless and on the contrary to the hydrogenated case, the Gaussian chain model does not reproduce well the experimental data and in particular, the amplitude of the oscillation. For the deuterated case, the fit with the core-shell model is better. The parameters deduced from the fit, thickness and polymer density of the grafted corona are of the same order of magnitude of the hydrogenated case. There is no significant difference between $d_1$ and $d_2$ even if this is expected regarding the different conversion rates between the two polymerizations.

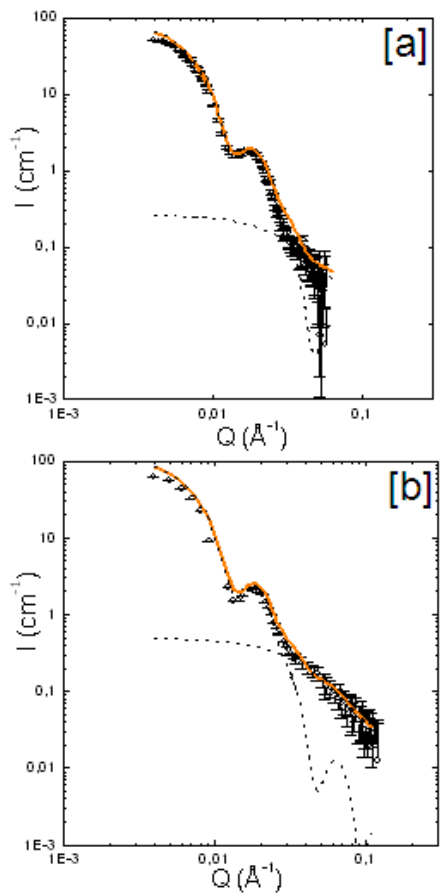

**Fig. 11** Scattering curve of a diluted solution of silica Ludox particles grafted with deuterated PS chains ($d_1$ [a], $d_2$ [b]) in contrast conditions for which the scattering contribution of the silica core is matched with a mixture of hydrogenated and deuterated solvent (DMAc). The full orange line is the result of the addition of the calculated core-shell model with the Debye simulation of a single chain which are presented in dash line (see text for details).

Although the core-shell model is better than the Gaussian chain model, the agreement is nevertheless not perfect for the whole Q range: the core-shell model decrease as $Q^{-4}$ at large Q and the experimental data as $Q^{-2}$. To account for this behavior, we have built a new model based on our analysis of the respective qualities of the core-shell and of the Gaussian chain model. The idea is to reproduce the curve by addition of the core-shell model which gives a good agreement at low and intermediate Q values and the behavior of a single grafted chain at large Q values which can be calculated with the Debye formula (equation 9). Note that such chain scattering is, at large Q, close to the one given by the Gaussian chain model. The radius of gyration and the concentration of the grafted chains have been deduced from the fit with the Gaussian chain model. The corresponding calculation is reported on figure 11 together with the best fit results for the core-shell model for $d_1$ (figure 11[a]) and $d_2$ (figure 11[b]). The sum of these two calculated scattering is reported as an orange full line. We can observe that the agreement with experimental data is now very consistent.

## Discussion

The first point of this work is the ability to obtain well-defined silica particles stabilized in an organic medium in which the polymerization of styrene can be efficient. We have been successful in transferring commercial Ludox particles from water into DMAc solvent while keeping the colloidal stability of the particles. The main issue of this transfer is to make now available particles with a well known form factor and a low polydispersity distribution. These informations are of a great importance for characterization of the objects after grafting. The second main improvement is to determine the parameters for the synthesis of polystyrene-grafted-silica nanoparticles in a simplified way (mono-component system) with efficient conversion rate and a good reproducibility. The grafted silica polystyrene particles must be produced in large quantities to be used later as fillers in polymer nanocomposites, the study of which requires many large samples. We chose a method based on Nitroxide-Mediated Polymerisation (NMP): the alkoxyamine which acts as initiator controller agent is bound to the silica nanoparticles surface in two steps. To make the best use of neutron contrast variation techniques, the monomer, styrene, was either normal or deuterated. With the help of model polymerization without silica particles, we could refine the synthesis parameters and obtain around 50% of conversion rate without any aggregation of the sol. The two steps of grafting give us high grafting density of first, the initiator and, second the grafted chains: a mean value of 400 chains/particle, close to 0.2 chain/nm² is satisfactory, as well as the chains molecular mass, around 30 000 Dalton. The whole set of results is consistent with good control and efficiency of the polymerization.

One of the still open questions concerning such polymerization process is the use of free initiator, which in our case also plays the role of control agent. Suppressing or reducing the free initiator could be of interest for varying the mass of the grafted chain but also to ascribe more precisely the radical persistent effect. Moreover and more practically, it could permit to skip the purification step required to remove the free synthesized chains. Our attempts without initiator showed a dramatical reduction of efficiency of the polymerization. Even if initiation, which becomes difficult in this case, starts, the initiator concentration is also too low to enable control: the chain mass is thus very polydisperse and the colloidal sol looses stability and becomes a gel. Another route of research is to add in the sol only the controller agent part of the initiator (the radical SG1) to ensure the control of the polymerization process while avoiding the creation of free chains. This way is currently under tests. Nevertheless, the synthesis parameters defined through this work, are well-optimized in term of conversion rate, reproducibility and colloidal stability to allow a refined characterization. This is, to our best knowledge, the first time that the controlled synthesis of nanoparticles grafted with deuterated polymer is reported in full details in literature.

Let us come to the characterization steps, mostly using SANS, which here are full part of the synthesis procedure. The first step is the analysis of the scattering curves from the grafted particles with matching of the polymer chains: it is crucial to check the aggregation state of the colloidal sol after the synthesis and the purification procedure. If this check is positive (no aggregation between the grafted particles), which is the case as we can see on the figure 6[a], it is possible to extract the inter-particles structure factor by dividing the total scattering by the silica form factor (figure 6[b]), and analyze the interactions between the particles in solution (which can still be measurable in spite of the absence of strong aggregation). Results show that a dilute solution of grafted particles, typically 0.5-1% v/v, present a nicely repulsive liquid-like order and thus does not present strong interaction between the particles; the structure factor $S(Q)$ is close to one over the full Q range. In our case, the use of scattering techniques in the reciprocal space is more suitable than a direct space microscopy method for the characterization of the form, the structure factor and the homogeneity of grafted particles as we extracted a mean representative value on a large number of atoms without to destabilized the interactions and with a better contrast. Such knowledge of the dispersion state of grafted particles is important to get into the modelization of the form factor of the grafted polymer corona. Is $S(Q)$ was different from one, we should divide the silica matched data by $S(Q)$ to get the form factor of the corona (as long as there is no complete aggregation leading to interpenetration of the coronas). The scattering from the grafted particles with matching of the silica core has been fitted with a basic core-shell model and with a Gaussian chain model according to a step-by-step fitting procedure. The fits have been applied to the two polymerizations repeated with same parameters with normal styrene ($h_1$, $h_2$) and to the two with deuterated styrene ($d_1$, $d_2$). Samples $h_1$, $h_2$, $d_1$ have been made in the same synthesis conditions while $d_2$ has been made with a lower monomer initial concentration (18% v/v instead of 30% v/v).

The first important result of this study is that the employed models reproduce the experimental scattering data along the whole Q range with a very good agreement: the particles grafted with hydrogenated chains ($h_1$, $h_2$) are well fitted to the Gaussian Chain model (figure 7[b] and 8[b]) while the curves of the particles grafted with the deuterated chains ($d_1$, $d_2$) are better fitted to the core-shell model (figure 9[a] and 10[a]). The deuterated corona could also be modelized very satisfactorily, in a second approach, with a combination of the core-shell model and a calculated scattering from polymer chains (Debye formula) which accounts for the $Q^{-2}$ dependence of the intensity in the large Q range (figure 11[a] and [b]). The second important result is the very good agreement between the chemical analysis and the scattering fitting results for the principal parameters which are the same for both fitting models: the particles volume fraction between 0.5 and 1% v/v, the size of the grafted chains Mw=30 000 g/mol corresponding to a radius of gyration of 60Å and the grafting density around 400 chains/particles.

Despite of these excellent agreements, the experimental data, and the two corresponding fitting models, exhibited some differences between the two labelling H/D. The transition regime between the $Q^{-4}$ decrease of the intensity and the $Q^{-2}$ regime at larger Q is actually one of the most striking differences: the $Q^{-2}$ appears sooner and is more visible for the hydrogenated case, while the $Q^{-4}$ regime extends at larger Q for the deuterated case. Therefore this behaviour can be related to the idea that the deuterated grafted layer is denser that the hydrogenated one. The results of the chemical analysis are in agreement with this assumption but the accuracy of such dosage is not enough for a safe conclusion. The question is whether it is physically relevant to link the scattering differences to a structural difference in the layer density or to an optical difference between solvent and normal or deuterated chains. Under the same synthesis conditions, there is no trivial evidence that the polymerization process will be different with a hydrogenated monomer and with a deuterated one. Nevertheless, in practice, the conditions of synthesis, purification, stabilization, purity degree and packaging can be different for the two isotopic species. This can influence the reactivity of the monomer at the surface of the nanoparticles during the synthesis and induce some variations in the final grafting densities. Since deuterated monomers are delivered in a distilled version, we tried to distil the hydrogenated monomer to obtain a monomer closer to the purity state of the deuterated one. The polymerization with distilled hydrogenated monomer does not give any change in comparison with the non distilled hydrogenated monomer case. However, other – yet unknown – isotopic differences in chemical reactivity cannot be excluded. We also must keep in mind physico-chemical aspects, like isotopic differences in surface adsorption, or in the solvent quality for polymer, which can be enhanced by the high chain concentration at the vicinity of the surface.

The second main explanation of the difference between the hydrogenated case and the deuterated case could be a contrast effect, suggested by the amplitude of the oscillation in the intermediate Q range. The two measured form factors depend directly on the square difference between the scattering length density of the chain and the scattering length density of the solvent. This difference is in a ratio 4 to 10 when passing from the hydrogenated to the deuterated case. In principle, if the d-PS and h-PS profiles are identical, the resulting scattering should be proportionnal. But the larger deuterated neutron contrast step lets appears an abrupt profile of the variation of the scattering length density. Such profile is close to the core-shell model, based on an abrupt variation of the radial profile of the scattering length density, as illustrated by the dominant $Q^{-4}$ regime in the intermediate Q range. On the contrary, a softer contrast step agrees better with a model with slow variation of the chain concentration radial profile. This is the case of the Gaussian model, as shown in the case of block copolymer, gives profiles consistent with power law behaviours as predicted by scaling theory[34, 38]. This soft profile result in a dominant $Q^{-2}$ regime in the intermediate Q range. Such "optical" explanation could be checked by the way of complementary experiments, in which we could investigate intermediate neutron contrast conditions (D/H solvent fractions in between silica matching and polymer matching ones) in order to follow the evolution of the fitting agreement between the experimental data and the two models as a function of the contrast.

## Summary and conclusions

We demonstrated the possibility to obtain well-defined silica nano-particles grafted with a regular normal or deuterated polystyrene corona. Starting from commercial silica particles in water, for the need of the synthesis, we successfully transferred them into an organic solvent, DMAc, without any destabilization of the colloidal equilibrium of the nano-particles. We proposed a convenient and efficient "grafting from" method based on Nitroxide-Mediated Polymerization via an original covalent grafting strategy of the alkoxyamine which acts both as initiator and controller agent. The synthesis parameters have been successfully optimized to obtain good reproducible conversion rates, and expected molecular weight of grafted chains, while avoiding the aggregation of the particles. After removing the free polymer chains (due to the presence of free initiator in volume), the grafted particles have been precisely characterized with SANS using neutron contrast matching. Matching the polymer chains permits to check the neat dispersion state of the silica particles at the end of the polymerization process. Matching the silica core allows to analyze quantitatively the signal of the grafted chains both for normal and deuterated polystyrene with two fitting models. These results show clearly that for each case the grafted chains form a regular corona around a single silica particle and a very good agreement between the results of the chemistry analysis and of the neutron fitting procedure for the parameters of the grafted chains: particles volume fraction, chain molecular weights and grafting density. Nevertheless, the form factor of the grafted chains seems to depend on the isotopic nature of the chains: whereas the normal grafted PS chains is well described with a Gaussian Chain Model, the deuterated grafted chains is better reproduced with a modified core-shell model (the sum of the core-shell model with a Debye chain). The origin of this observation can be either analytical, the solvent-layer neutronic amplitude contrast being related to the calculated form factor used for fitting, or structural, the deuterated layer being denser due to higher monomer chemical reactivity or isotopic dependences of the physical-chemistry. In such case this mean that the radial polymer concentration profiles remains to be refined, but this means also that SANS is a very powerful and refined method to investigate these details. To conclude, at the stage of this paper, SANS essentially appears to be a safe characterization method since all our models give the same principal parameters: individual dispersion, total molecular weight, global size, core size which guarantee the existence of a polymer layer around each particle. Hence, the synthesis and the characterization processes described in this work constituted a robust combination for the production of well defined grafted nanoparticles which will be used as filler in nano-composites for further reinforcement studies.

## Acknowledgments

We thank Jérôme Vinas for his help on the chemistry process, in particular his work on the grafting of SG1 initiator and Alain Lapp for his friendly help on PAXY. We thank ARKEMA for providing BlocBuilder®, CEA, CNRS and Université de Provence for financial support.